# Beyond the Shell Model: The Canonical Nuclear Many-Body Problem as an Effective Theory

W. C. Haxton and C.-L. Song

*Institute for Nuclear Theory, Box 351550, and Department of Physics, Box 351560,*
*University of Washington, Seattle, WA 98195*

(June 12, 2017)

We describe a strategy for attacking the canonical nuclear structure problem —bound-state properties of a system of point nucleons interacting via a two-body potential—which involves an expansion in the number of particles scattering at high momenta, but is otherwise exact. The required self-consistent solutions of the Bloch-Horowitz equation for effective interactions and operators are obtained by an efficient Green's function method based on the Lanczos algorithm. We carry out this program for the simplest nuclei, d and $^3$He, to contrast a rigorous effective theory with the shell model, thereby illustrating several of the uncontrolled approximations in the latter.

In this letter we argue that it may be possible to move beyond the nuclear shell model (SM) to a more rigorous treatment of the canonical nuclear structure problem of $A$ nonrelativistic point nucleons interacting via a two-body potential. Our optimism is inspired by several recent developments. One is the success of the Argonne group's efforts [1] to predict the properties of light nuclei in effectively exact Green's function Monte Carlo calculations, using an NN potential carefully fit to scattering data and augmented by weaker three-body forces. This suggests that SM failures have their origin in an incomplete treatment of the many-body physics, rather than in the starting Hamiltonian. A second is the success of effective field theory (EFT) treatments of the two- and three-body problems. This work not only provides some insight into why such a starting Hamiltonian is reasonable, but has made the community more aware of the uncontrolled approximations implicit in the SM and other approaches.

The SM's strength is its explicit representation of $\sim$ 60% of the wave function that resides at long wavelengths: the $A$-body correlations important to collective modes are addressed by direct diagonalization. A third development is the remarkable recent advances in such SM technology, including Lanczos-based methods [2], treatments of light nuclei involving many shells [3], and Monte Carlo sampling algorithms [4,5]. The dimensions of tractable SM spaces have risen by several orders of magnitude in the past few years.

Such diagonalizations in a long-wavelength "included space" could be an important piece of a rigorous effective theory (ET) of nuclear structure in which the Hamiltonian operating in an infinite Hilbert space

$$H = \frac{1}{2}\sum_{i,j=1}^{A}(T_{ij}+V_{ij}) \rightarrow \frac{1}{2}\sum_{i,j,...=1}^{A} H^{eff}_{ij...} \quad (1)$$

is replaced by an $H^{eff}$ operating in a finite SM space. The effects of high momentum components appear as effective contributions to the Hamiltonian and operators. The hope in nuclear physics, inspired by Brueckner's treatment of nuclear matter, is that the "excluded space" integration might be carried out as a rapidly converging series in the number of nucleons scattering at one time in high momentum states. In this way the effective theory might prove far more tractable than the original $A$-body problem in a infinite Hilbert space.

Despite some text book motivations, the SM is a model rather than an ET:

1) The functional form of the SM effective interaction, $\langle\alpha\beta|H^{eff}|\gamma\delta\rangle$, is correct only in lowest order and only if the calculation is restricted to a single shell [6]. In a faithful ET three-, four-, and higher-body operators are successively added, and the matrix elements generally carry, in addition to single-particle quantum numbers, an index specifying the number of quanta carried by the remaining spectator nucleons.

2) Typically $H^{eff}$ lacks the symmetries of the original bare $H$, e.g., translational invariance and hermiticity (though the latter is often enforced by hand).

3) SM wave functions are orthogonal and normed to unity. In ET the effective wave functions are naturally defined as the restrictions of the true wave functions $|\Psi_i\rangle$ to the model space, so that the norms are less than unity and orthogonality is lost.

4) Shell model interactions frequently depend on fictitious parameters such as "starting energies," introduced to adjust the energy denominator in the two-body G-matrix or to account for intermediate-state average energies when the two-body G-matrix is iterated to produce some approximation to a higher order $H^{eff}$.

5) Perhaps most serious, the important issue of effective operators is almost never addressed in a meaningful way. In most cases SM $H^{eff}$s are derived phenomenologically, so that there is no diagrammatic basis for generating the effective operator. Thus empirical operator renormalizations must also be introduced, obscuring the underlying physics and undercutting the SM as a predictive tool.

In this letter we describe the first steps in an effort to assess the feasibility of an exact ET "SM"-like theory. The approach is sketched in Fig. 1. The Hilbert space is divided into a long-wavelength "SM" space, defined by some energy scale $\Lambda_{SM}$, and a high-momentum space. One can truncate the latter at some scale $\Lambda_\infty \sim 3$ GeV, characteristic of the cores of realistic potentials, as above this energy, excitations make a negligible contribution. All correlations within the "SM" space are included, but the high-momentum correlations in the excluded space are limited to $n$-body, where $n$ is the cluster size. Thus Figs. 1b and 1c give the lowest and next-to-lowest approximations to $H^{eff}$. Note that one can view Fig. 1c as containing Fig. 1b, a density-dependent two-body correction to Fig. 1b, and true 3-body terms. The pattern continues as $n$ is increased: true $n$-body terms are introduced and all lower-order results are corrected to one higher order in the density $\rho$.

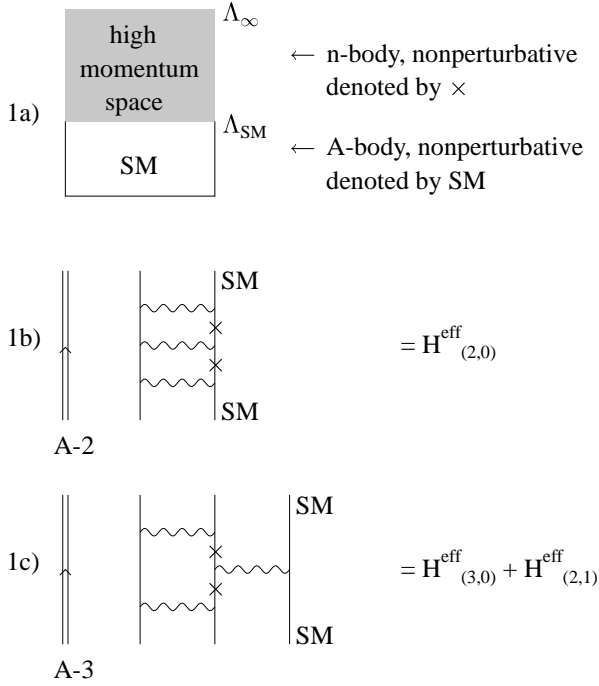

FIG. 1. Cluster expansion of the effective interaction.

If the "SM" space is defined as all harmonic oscillator Slater determinants with $E \leq \Lambda_{SM}$, $H^{eff}$ becomes translationally invariant and the ladder sums can be carried out in relative coordinates, a considerable simplification. The projection operator onto the high-momentum space $Q$ thus depends on $\Lambda_{SM}$ and the oscillator parameter $b$, where the latter can be chosen to optimize the convergence in $\Lambda_\infty$ [7].

The resulting Bloch-Horowitz (BH) equation [8] is then

$$H^{eff} = H + H \frac{1}{E - QH} QH$$
$$H^{eff}|\Psi_{SM}\rangle = E|\Psi_{SM}\rangle \quad |\Psi_{SM}\rangle = (1-Q)|\Psi\rangle \quad (2)$$

where $|\Psi\rangle$ is the exact wave function and $H|\Psi\rangle = E|\Psi\rangle$. These equations must be solved self-consistently because $H^{eff}$ depends on the unknown eigenvalue E. The harmonic oscillator appears only implicitly through $Q$ in distinguishing the long-wavelength "SM" space from the remainder of the Hilbert space.

There is an extensive literature on this and similar equations [9,10]. Frequently $H$ is divided into an unperturbed $H_0$ and an perturbation $H - H_0$, but well-known pathologies due to intruder states can affect the resulting perturbation expansion [11]. The approach explored here is nonperturbative and uses the Lanczos algorithm to sum the $n$-body ladders.

The Lanczos algorithm recursively maps a hermitian operator $H$ of dimension $N$ into tridiagonal form

$$\begin{aligned} H|v_1\rangle &= \alpha_1|v_1\rangle + \beta_1|v_2\rangle \\ H|v_2\rangle &= \beta_1|v_1\rangle + \alpha_2|v_2\rangle + \beta_2|v_3\rangle \\ H|v_3\rangle &= \quad\quad \beta_2|v_2\rangle + \alpha_3|v_3\rangle + \beta_3|v_4\rangle \;\; etc. \end{aligned} \quad (3)$$

If this process is truncated after $n \ll N$ steps, the resulting matrix contains the information needed to reconstruct the exact $2n$-1 lowest moments of $|v_1\rangle$ over the eigenspectrum. One of the applications of this algorithm is in constructing fully interacting Green's functions [12] as a function of $E$,

$$\frac{1}{E-H}|v_1\rangle = g_1(E)|v_1\rangle + g_2(E)|v_2\rangle + \cdots \quad (4)$$

where the $g_i(E)$ are continued fractions that depend on $\alpha_i, \beta_i$ and where E appears only as a parameter.

Thus a simple procedure can be followed to solve the BH equation:
• For each relative-coordinate vector in the SM space $|\gamma\rangle$, form the excluded-space vector $|v_1\rangle \equiv QH|\gamma\rangle$ and the corresponding Lanczos matrix for the operator $QH$. Retaining the resulting coefficients $\alpha_i, \beta_i$ for later use, construct the Green's function for some initial guess for $E$ and then the dot product with $\langle \gamma'|H$ to find $\langle \gamma'|H^{eff}(E)|\gamma\rangle$.
• Perform the "SM" calculation to find the desired eigenvalue $E'$ which, in general, will be different from the guess $E$. Using the stored $\alpha_i, \beta_i$, recalculate the Green's function for $E'$ and $H^{eff}(E')$ then redo the "SM" calculation. The process is repeated until the energy is fully converged.
• Then proceed to the next desired bound state and repeat the process. Note that it is not necessary to repeat the $H^{eff}$ calculation. The eigenvalue taken from the "SM" calculation is, of course, that of the next desired state, yielding a distinct $H^{eff}$ for each eigenvalue.

The attractiveness of this approach is that the effective interactions part of the procedure, which is relatively time consuming as it requires one to perform a large-basis Lanczos calculation for each relative-coordinate starting vector in the "SM" space, is performed only once. While the SM calculation must be repeated in the iterations, the convergence is rapid (6-8 cycles, typically). As modern workstations can manage calculations of dimension $10^6$ in about 30 minutes, the overall procedure is certainly practical.

The technical aspects of this approach are described elsewhere [7,13]. Here we focus on the results for the simplest nuclei, d and $^3$H, carrying the above process to completion (two- and three-body ladders, respectively).

The binding energies and operator matrix elements for simple systems like $^3$He can, of course, be calculated exactly by other methods. The point of our work is not to offer an alternative to these techniques for these nuclei, but rather to illustrate the conceptual differences between a faithful ET and the SM. We performed d and $^3$He ET calculations for a series of "SM" spaces (2, 4, 6, and 8 $\hbar\omega$), in each case using the Lanczos Green's function algorithm to evaluate the two- and three-body ladders (100 Lanczos iterations are more than sufficient) and iterating the "SM" calculation until the results are fully converged. The deuteron calculation is rather trivial; for $\Lambda_\infty \sim 50$ the $^3$He calculation involves a dense matrix of $\sim 10^4$, still quite modest by current SM standards. (The Hamiltonian matrix is dense because relative Jacobi coordinates are used, rather than the m-scheme, together with standard Talmi-Brody-Moshinsky methods [13,14].) A first test of this procedure is the stability of the resulting energy eigenvalue: the results for the four chosen "SM" spaces agreed to four places, -2.224 MeV (using $\sqrt{2}b = 1.6f$ and $\Lambda_\infty = 140$). The exact result is -2.2246 MeV. The $^3$He results were similarly very stable.

A more interesting test is the evolution of the wave function as the "SM" space is enlarged. Table I gives results for $^3$He. (The procedure for calculating the wave function normalization is discussed below.) Unlike typical SM calculations, the amplitudes agree over overlapping pieces of the Hilbert spaces. As one proceeds through $2\hbar\omega$, $4\hbar\omega$, $6\hbar\omega$, ... calculations, the ET wave function evolves only by adding new components in the expanded space. Consequently, as Table I shows, the wave function norm grows.

This evolution will not arise in the standard SM because the wave function normalization is set to unity regardless of the model space. It will also not arise for a second reason, illustrated in Table II. The three-body $^3$He matrix elements of $H^{eff}$ are crucially dependent on the model space: a typical matrix $\langle \alpha|H^{eff}|\beta \rangle$ changes very rapidly under modest expansions of the model space, e.g., from $2\hbar\omega$ to $4\hbar\omega$. Yet it is common practice in the SM to expand calculations by simply adding to an existing SM Hamiltonian new interactions that will mix in additional shells. We suspect the behavior found for $^3$He is generic in ET calculations: it arises because a substantial fraction of the wave function lies near but outside the model space (e.g., see Table I). An expansion of the model space changes the energy denominators for coupling to some of these configurations, and moves other nearby configurations from the excluded space to the model space. Naively, relative changes in effective interaction matrix elements of unity are expected.

TABLE I. ET results for the $^3$He ground state wave function calculated with the Argonne $v18$ potential. Selected basis states are designated somewhat schematically as $|N,\alpha\rangle$, where $N$ is the total number of oscillator quanta and $\alpha$ is an index representing all other quantum numbers.

| state | amplitude | | | | | |
|---|---|---|---|---|---|---|
| | $0\hbar\omega$ | $2\hbar\omega$ | $4\hbar\omega$ | $6\hbar\omega$ | $8\hbar\omega$ | exact |
| | (31.1%) | (57.4%) | (70.0%) | (79.8%) | (85.5%) | (100%) |
| $\|0,1\rangle$ | 0.55791 | 0.55791 | 0.55791 | 0.55795 | 0.55791 | 0.55793 |
| $\|2,1\rangle$ | 0.00000 | 0.04631 | 0.04613 | 0.04618 | 0.04622 | 0.04631 |
| $\|2,2\rangle$ | 0.00000 | -0.48255 | -0.48237 | -0.48243 | -0.48243 | -0.48257 |
| $\|2,3\rangle$ | 0.00000 | 0.00729 | 0.00731 | 0.00730 | 0.00729 | 0.00729 |
| $\|4,1\rangle$ | 0.00000 | 0.00000 | -0.02040 | -0.02042 | -0.02043 | -0.02047 |
| $\|4,2\rangle$ | 0.00000 | 0.00000 | 0.11267 | 0.11274 | 0.11275 | 0.11289 |
| $\|4,3\rangle$ | 0.00000 | 0.00000 | -0.04191 | -0.04199 | -0.04208 | -0.04228 |

TABLE II. Selected BH 3-body effective interaction matrix elements for $^3$He, in MeV, illustrating the strong dependence on the "SM" space.

| | $2\hbar\omega$ | $4\hbar\omega$ | $6\hbar\omega$ | $8\hbar\omega$ |
|---|---|---|---|---|
| $\langle 0,1 \| H^{eff} \| 2,1 \rangle$ | -4.874 | -3.165 | -0.449 | 1.279 |
| $\langle 0,1 \| H^{eff} \| 2,5 \rangle$ | -0.897 | -1.590 | -1.893 | -2.208 |
| $\langle 2,1 \| H^{eff} \| 2,2 \rangle$ | 6.548 | -2.534 | -4.144 | -5.060 |

Now we turn to the question of operators. The standard procedure in the SM is to calculate nuclear form factors with bare operators, or perhaps with bare operators renormalized according to effective charges determined phenomenologically at $q^2 = 0$, using SM wave functions normed to 1. As we now have a series of exact effective interactions corresponding to different model spaces, we can test the validity of this approach. The results for the elastic magnetic form factors are given in Fig. 2. Even though each "SM" $H^{eff}$ is, in a sense, perfect, the results for bare operators are widely divergent at even modest momentum transfers of $\sim 2/f$. This is not surprising: if an operator transfers a momentum $q \gtrsim 2k_F$ to the nucleus, where $k_F$ is the Fermi momentum, the resulting amplitude should reside primarily outside the "SM" space, where it contributes only through effective pieces of the operator.

Clearly the effective interaction and effective operator have to be treated consistently and on the same footing. The bare operator $\hat{O}$ must be replaced by

$$\hat{O}^{eff} = (1 + HQ\frac{1}{E_f - HQ})\hat{O}(1 + \frac{1}{E_i - QH}QH) \quad (5)$$

and must be evaluated between "SM" wave functions

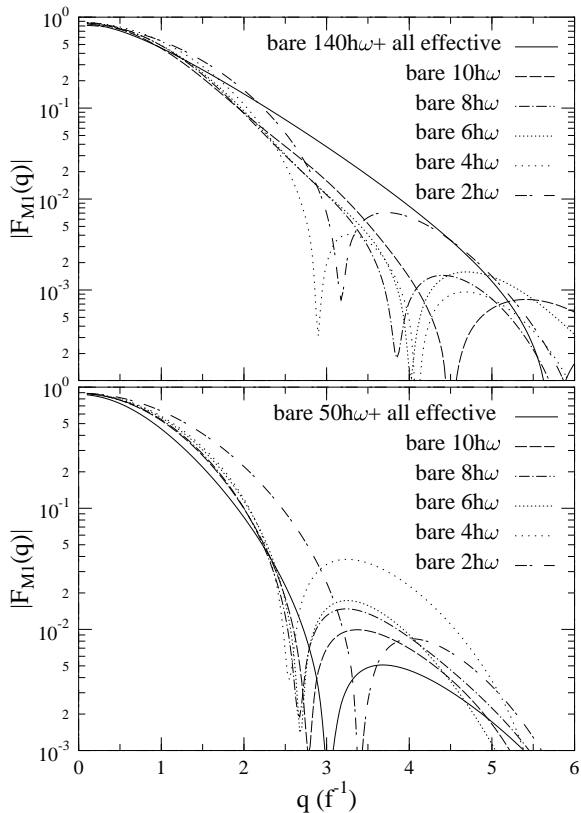

FIG. 2. The magnetic elastic form factors for the deuteron (top) and $^3$He (bottom) calculated with the exact $H^{eff}$, SM wave functions normalized to unity, and a bare operator are compared to the exact result (solid line). When effective operators and the proper wave function normalizations are used, all results become identical to the solid line.

normed according to

$$1 = \langle \Psi_i | \Psi_i \rangle = \langle \Psi_i^{\text{``SM''}} | \hat{1}^{eff} | \Psi_i^{\text{``SM''}} \rangle \qquad (6)$$

(and similarly for $|\Psi_f^{\text{``SM''}}\rangle$). These expressions can be evaluated with the Lanczos Green's function methods described earlier. When this is done, all of the effective calculations, regardless of the choice of the model space, yield the same result, given by the solid lines in Fig. 2.

It is likely that many persistent problems in nuclear physics — such as the renormalization of $g_A$ in $\beta$ decay — are due to naive treatments of operators. It should be apparent from the above example that no amount of work on $H^{eff}$ will help with this problem. What is necessary is a diagrammatic basis for generating $H^{eff}$ that can be applied in exactly the same way to evolving $\hat{O}^{eff}$. From this perspective, phenomenological derivations of $H^{eff}$ by fitting binding energies and other static properties of nuclei are not terribly helpful, unless one intends to simultaneously find phenomenological renormalizations for each desired operator in each $q^2$ range of interest.

There are several remarks we would like to make in conclusion. First, the demonstration that one can efficiently solve the three-body problem as an exact ET in a SM-like model space is already quite significant: this means it is relatively straightforward to execute a faithful BH treatment of heavier nuclei through order $\rho$ in both the effective interaction and effective operators. The numerical effort is comparable. Second, our results indicate the many of the approximations in the SM are uncontrolled, so that a more faithful ET is badly needed. Third, further progress can be made numerically: we expect to gain a factor of 50 when the large-memory capabilities of our workstation are properly exploited. Finally, there is the strong possibility of theoretical improvements. For example, the numerical convergence of our binding energies for large $\Lambda_\infty$ is exponential, $\exp(-a\Lambda_\infty^2)$. We argue in [7] that the origin of this scaling is the contraction of ladders at very high excitation energies to local operators, which can be summed analytically in a local density expansion. If this conjecture proves true, we can bring $\Lambda_\infty$ down to a much lower scale, with no loss of precision. Thus much remains to be done.

This work was supported in part by the Division of Nuclear Physics, US Department of Energy. We thank David Kaplan, Peter Lepage, and Martin Savage for helpful comments.